\documentclass{aastex631}
\usepackage{epsfig,amsmath,amsfonts,amssymb,graphicx,subfigure,enumitem,color}
\usepackage{amssymb}
\usepackage{amsmath}
\usepackage[varg]{txfonts}
\usepackage{natbib}
\usepackage{url}             % For breaking URLs easily through lines
\usepackage{graphicx}
\usepackage{float}
\newcommand{\beq}{\begin{equation}}
\newcommand{\eeq}{\end{equation}}
\newcommand{\beqa}{\begin{eqnarray}}
\newcommand{\eeqa}{\end{eqnarray}}
\newcommand{\beqann}{\begin{eqnarray*}}
\newcommand{\eeqann}{\end{eqnarray*}}
\bibpunct{(}{)}{;}{a}{}{,} % to follow the A&A style

\usepackage{textcomp}
\usepackage{comment}

\shorttitle{Oscillations in quiet-Sun}
\shortauthors{Sangal et al.}
\graphicspath{{./}{figures/}}

\begin{document}

\title{Statistical investigation of wave propagation in the quiet-Sun using IRIS spectroscopic observations}

\correspondingauthor{A.K. Srivastava, Ding Yuan}
\email{asrivastava.app@iitbhu.ac.in, yuanding@hit.edu.cn}

\author{Kartika Sangal}
\affiliation{Department of Physics, Indian Institute of Technology (BHU), Varanasi-221005, India} 
%\\
%1667 K Street NW, Suite 800 \\
%Washington, DC 20006, USA}

\author{A.K.~Srivastava}
\affiliation{Department of Physics, Indian Institute of Technology (BHU), Varanasi-221005, India}

%\collaboration{6}{(AAS Journals Data Editors)}

\author{P.~Kayshap}
\affiliation{Vellore Institute of Technology, Kotri Kalan, Ashta, Near, Indore Road, Bhopal, Madhya Pradesh 466114}
%\affiliation{AAS Journals Associate Editor-in-Chief}

\author{Ding~Yuan}
\affiliation{Institute of Space Science and Applied Technology, Harbin Institute of Technology, 518055, Shenzhen, Guangdong, China}
\affiliation{Shenzhen Key Laboratory of Numerical Prediction for Space Storm, Harbin Institute of Technology, Shenzhen, Guangdong 518055, China}
\affiliation{Key Laboratory of Solar Activity and Space Weather, National Space Science Center, Chinese Academy of Sciences, Beijing 100190, China}

\author{E.~Scullion}
\affiliation{Department of Mathematics \& Information Sciences,Northumbria University, Newcastle Upon Tyne, NE1 8ST, UK}

%% Mark off the abstract in the ``abstract'' environment. 
\begin{abstract}

In the current analysis, we use spectroscopic observations of the quiet-Sun made by  IRIS instrument, and investigate wave propagation.
We analyze various  spectral lines formed in different atmospheric layers such as the photosphere, chromosphere, and transition region. We examine Doppler velocity time-series at various locations in the quiet-Sun to determine the dominant oscillation periods. Our results 
executing statistical analysis resemble those of the classical physical scenario, indicating that the photosphere is mainly characterized by the dominant 5-minute period, while the chromosphere is primarily associated with the 3-minute oscillation period. In the transition region, we observe a variety of oscillation periods, with dominant periods of 3, 8, and 12 minutes. We estimate the cut-off frequency by deducing  phase difference between two Doppler velocity time-series obtained from spectral line pairs in different atmospheric layers formed at different temperatures. It reveals a significant correlation between 3-minute periods in TR and photospheric oscillations, suggesting that these oscillations in the TR might propagate from the photosphere. Additionally, we analyze the phase difference between chromospheric oscillations and photospheric oscillations, demonstrating that only the 3-minute oscillations propagate upwards. Based on the statistical analyses, we suggest the presence of magnetoacoustic waves in the solar atmosphere in which some are propagating from the lower solar atmosphere upward, while some others are propagating downward. TR carries both long-period oscillations generated {\it in situ}, and some photospheric oscillations which are also able to reach there from below.

\end{abstract}


\section{Introduction} \label{sec:intro}

The acoustic cutoff frequency was introduced by \cite{10.1112/plms/s2-7.1.122} as the ratio between the sound speed and twice the pressure or density scale height. His definition was based on the assumption of an isothermal atmosphere, and hence the resulting cutoff frequency was a constant global value throughout the solar atmosphere. Later researchers have extended Lamb's work and put forth alternative equations for calculating the acoustic cutoff frequency in the solar atmosphere with temperature gradients \citep[e.g.,][and references therein]{1984ARA&A..22..593D,1998A&A...337..487S,2006PhRvE..73c6612M,2012MNRAS.421..159F,2014AN....335.1043R}. Recent studies have been directed us on examining how the cutoff frequency varies with height in the solar atmosphere, rather than assuming it to be a global quantity \citep[e.g.,][]{2016ApJ...819L..23W,2016ApJ...827...37M,2019A&A...627A.169F}.

Understanding the existence of cutoff frequency or period that enable acoustic waves to propagate in the solar atmosphere is crucial for comprehending the source of chromospheric oscillations. \cite{2009ApJ...692.1211C} conducted a study on the cutoff frequency and wave propagation properties, with respect to the magnetic flux of different solar structures (e.g., including two sunspots of different sizes, a tiny pore, and a facular region). The results of these studies have consistently shown that the energy reaching the upper chromosphere beyond the acoustic cutoff frequency originated exclusively from the photosphere in all the observed cases. \cite{2000ApJ...531.1150W} studied the oscillations in the spectral time series in chromosphere/lower transition region above the internetwork area. They have found that periods between 120 and 200 s are dominant oscillatory periods. With the help of phase difference analysis, they have shown drivers of these oscillations as p-mode oscillations.

The acoustic cut-off frequency or period is a critical parameter in the solar atmosphere, as it determines the propagation threshold of acoustic waves from the photosphere to the upper atmosphere. Consequently, it determines the propagation conditions for the acoustic waves \citep[e.g.,][]{10.1112/plms/s2-7.1.122,1964ApJ...139...48M,1993A&A...273..671F}. Waves with frequency lower than the acoustic cut-off frequency or a period greater than the acoustic cut-off period are unable to propagate into the upper atmosphere, and they become evanescent. On the other hand, the waves with a frequency higher than the acoustic cut-off frequency or a period less than the acoustic cut-off period can propagate into the upper atmosphere \citep[e.g.,][]{1979ApJ...231..570L, 1992ApJ...397L..59C}.

The long period waves (greater than the cut-off period) have been continuously observed in the solar atmosphere, thus enabling the debate of their existence. For example, in the chromospheric network region, long-period oscillations have been observed in various studies \citep[e.g.,][]{2007A&A...461L...1V,2011A&A...533A.116Y,2014MNRAS.443.1267B}. These oscillations can be influenced by the strong magnetic fields, which may alter the radiative relaxation time and lead to an increment in the cut-off period \citep[e.g.,][]{1983SoPh...87...77R,2002ApJ...567L.165M,2003ApJ...587..806M,2004ApJ...604..936B,2006ApJ...640.1153C,2008ApJ...676L..85K,2009ApJ...692.1211C,2010MNRAS.405.2317S}. Radiative losses is known to increase the cut-off period \citep[e.g.,][]{1983SoPh...87...77R,2008ApJ...676L..85K,2020A&A...640A...4F}. When the radiative relaxation time is adequately small, as is common in small-scale magnetic structures, it permits an increment of the cut-off period, thereby facilitating the propagation of 5-minute evanescent waves. In the numerical study done by \cite{2008ApJ...676L..85K}, they have shown a shift of the dominant period from 5 min to 3 min from the photosphere to the chromosphere in the case of adiabatic conditions. Conversely, there is no such transition in the dominant period when the radiative relaxation time is low. They subsequently compared power spectra with observational data and found that instances with shorter relaxation times resemble those observed in spectropolarimetric data of facular regions \citep{2006ASPC..358..465C}. Thus, they affirmed the significant impact of radiative losses on small-scale magnetic structures, such as those found in facular regions, capable of increasing the cut-off period and altering atmospheric transmission properties.
Furthermore, the inclination of the magnetic field can also increase the propagation of long-period waves \citep[e.g.,][]{2011ApJ...743..142H, 2004Natur.430..536D,2006RSPTA.364..383D,2006ApJ...647L..77M,2013ApJ...779..168J,2014A&A...567A..62K,2018MNRAS.479.5512K}. As magnetoacoustic waves align with a preferred direction set by the magnetic field, their effective cutoff frequency reduces in accordance with the cosine of the inclination angle. This alignment contributes to the waves propagation into the upper atmosphere. For instance, \cite{2004Natur.430..536D} propose that the magnetic field inclination is crucial for the escape of p-modes potent enough to generate the dynamic jets observed in active region fibrils. Following a similar train of thought, \cite{2006ApJ...648L.151J} assert that inclined flux tubes explain the propagating waves observed in the lower chromosphere. Theoretically, \cite{1977A&A....55..239B}  suggested that the cut-off period remains independent of the magnetic field in high $\beta$ plasma regions, while in low $\beta$ regimes, its value is influenced by the cosine of the angle $\phi$ formed by the magnetic field lines concerning the solar local vertical.

Previous studies have suggested that p-mode oscillations may be responsible for driving chromospheric 3 min oscillations \citep[e.g.,][and references therein]{2004ApJ...602..436M,2017ApJ...836...18C,2019A&A...627A.169F,2020A&A...642A.231L}. These oscillations are mainly influenced by magnetic field before they even reach the solar transition region. If a magnetic field is present, longer period oscillations ($>$ 4 min periodicities) can easily propagate into the transition region from the photosphere. \cite{2007A&A...461L...1V} conducted a study on oscillations in the quiet solar chromosphere and suggested that a considerable fraction of the acoustic power in the chromosphere, particularly at the frequencies below the acoustic cut-off and near the magnetic network elements, is transmitted directly from the photosphere. Hence, the network magnetic elements can act as a channel for the low-frequency photospheric oscillations to enter in the chromosphere, and therefore leading to the mechanical energy input to the upper atmosphere \citep{2003ApJ...595L..63D,2005ApJ...624L..61D,2008A&A...481L..95S}.

Our objective is to analyze the spectroscopic data ( i.e., multiple spectral lines) obtained from the IRIS instrument to investigate the oscillations %in the region %above the inter-network area 
in the quiet-Sun. Our analysis yields the determination of the dominant period of oscillations observed in different spectral lines formed in different atmospheric layers characterizing a certain formation temperature covering a wide range of the solar atmosphere from near the photosphere to the transition region.%at distinct altitudes across the solar atmosphere. 
~Moreover, we utilize cross-wavelet analysis and phase difference estimation to calculate the cutoff frequency (or cut-off period) between two different altitudes/regions in the solar atmosphere above the chosen patch of the quiet-Sun. In the present work, the diverse physical scenario of the wave propagation is reported using the statistical analysis of the estimated parameters (e.g., distribution of the statistically significant oscillatory periods, phase differences, cut-off frequency etc.) derived from observations of velocity time series. The paper is structured as follows in order to present our scientific findings. Section \ref{sec:data_mainbody} illustrates the observations, data analysis, and methodology, while Section \ref{sec:maximum_oscillatory_powers_TR} describe the results. Finally, in Section \ref{sec:discussion}, we outline discussion and conclusions. 

%%%%%%%%%%%%%%%%%%%%%%%%%%%%%%%%%%%%%%%%%%%%%%%%%%%%%%%%%%%%%%%%%%%%%%%%%%%%%%%%%%%%%%%%%%%%%%%%%%%%%%%%%%%%%%%%%%%%%%%%%%%%%%%%%%%%%%%%%%%%%%%%%%%%%%%%%%%%%%%%%%%%%%%%%%%%%%%%%%%%%%%%%%%%%%%%%%%%%%%%%%%%
\begin{figure*}
\centering
\includegraphics[width=.95\linewidth]{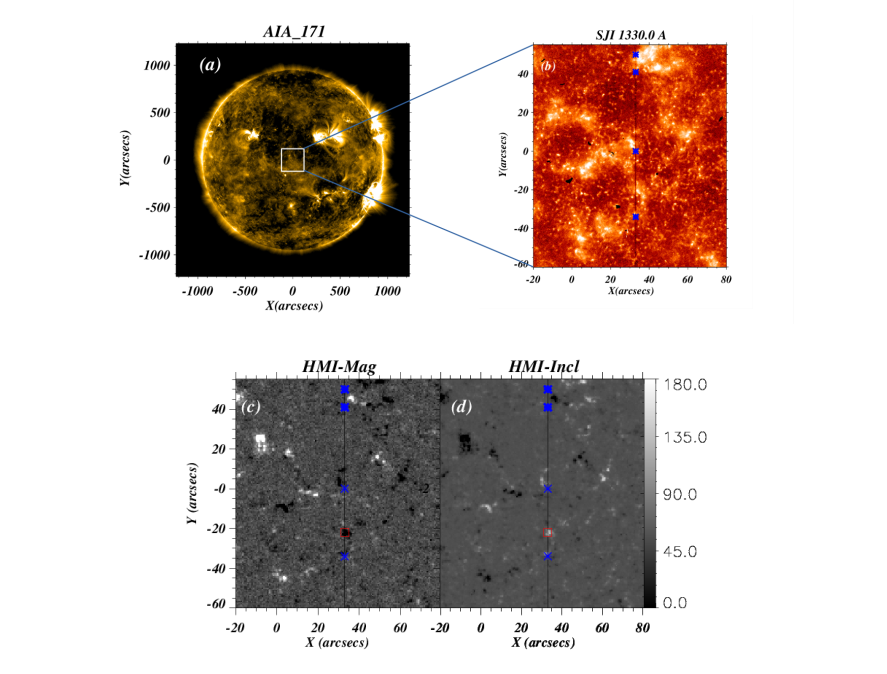}
    
 \caption{The panel (a) displays a full disc image of the Sun obtained by SDO/AIA at a wavelength of 171 \AA. The image has a white box superimposed on it to mark the region of interest. The panel (b) shows the zoomed-in view corresponding to the IRIS SJI 1330 \AA\ filter. The vertical black line is indicating the position of the slit used in our analysis. Our investigation is focused on the inter-network areas located between the blue asterisks. The panel (c) illustrates the absolute line-of-sight magnetic field, while panel (d) depicts the map of magnetic field inclination obtained by the HMI onboard the SDO. The magnetic field data is used as a context data to identify the physical nature of the region-of-interest in the quiet-Sun. The selected area corresponds to an inter-network region characterized by the low magnetic field strengths and high inclination angles, reaching up to 90\textdegree\ in the base-line of the HMI contextual observational data.}
    \label{fig:fig1}
\end{figure*}

%%%%%%%%%%%%%%%%%%%%%%%%%%%%%%%%%%%%%%%%%%%%%%%%%%%%%%%%%%%%%%%%%%%%%%%%%%%%%%%%%%%%%%%%%%%%%%%%%%%%%%%%%%%%%%%%%%%%%%%%%%%%%%%%%%%%%%%%%%%%%%%%%%%%%%%%%%%%%%%%%%%%%%%%%%%%%%%%%%%%%%%%%%%%%%%%%%%%%%%%%%%
\section{Analysis of The Observational Data} \label{sec:data_mainbody}

\subsection{Data Reduction}
\label{sec:data}

We analysed spectroscopic data from IRIS, which provides spectra with high resolution in the near-ultraviolet (NUV, 2783-2834 \AA) and ultraviolet (FUV, 1332-1358, 1389- 1407 \AA) regions \citep[IRIS;][]{2014SoPh..289.2733D,2021SoPh..296...84D}. The 0.33\arcsec wide IRIS slit can be used in either sit \& stare or raster scan mode. Additionally, IRIS offers to capture slit-jaw images (SJIs) with a 0.166\arcsec pixel scale at four different filters centered on 1400 \AA, 1330 \AA, 2796 \AA, and 2832 \AA. On July 5, 2021, IRIS observed a quiet-Sun region from approximately 16:59 to 22:57 UT. With a field of view of 0.33 x 119 $arcsec^2$ and a raster cadence of 17 seconds, the spectra are recorded in a sit and stare mode with the slit centered at the disc center (-1\arcsec,0\arcsec). This observation is well-suited for conducting multi-wavelength studies because it enables the temporal sampling of the same region with different lines that form at varying atmospheric heights.

The data used are the calibrated level 2 data. These include wavelength correction, geometry adjustments, and subtraction of the dark current. Figure \ref{fig:fig1}(a) depicts a context image of the full solar disc captured by the AIA filter 171 \AA. The white box overplotted on the image represents the region of interest (ROI) in the quiet-Sun, and zoomed-in view of the ROI in the SJI filter 1330 \AA\ is shown in the panel (b). The slit position on the SJI filter is represented by the black line. The spectra related to several emission and absorption lines are captured by the slit. The instrument is capable of recording the solar spectrum in both near-ultraviolet and far-ultraviolet regions, encompassing various photospheric, chromospheric, and transition region lines such as Ni I 2805.904 \AA~, Fe I 2799.972 \AA~, Mg II h 2802.704 \AA~, and Si IV 1402.77 \AA~. We have computed the presence of dominant oscillations and the propagation of waves at different altitudes using different spectral lines, such as Fe I, Mn I, Mg II h, and Si IV, which are exclusively formed at different altitudes (or atmospheric layers). Moreover, this particular observation is situated at the solar disk's center, where projection effects are less likely. Consequently, different lines that predominantly form at distinct atmospheric altitudes provide samples of the same region in the solar atmosphere at those altitudes. Moreover, we aim to make a statistical analysis of the presence of significant and dominant oscillatory periods in different layers ($\it viz.$ photosphere-chromosphere-TR) of the solar atmosphere, and thus physical behaviour of associated wave propagation. On the contrary, we certainly do not stress  on spatially resolving their presence on each and every location. Although, we utilize different chosen $Y-$positions of the IRIS slit to derive time series of the velocity using different spectra.

Distinguishing between two distinct magnetic regions is of utmost importance when we examine the quiet chromosphere. The magnetic networks are situated at the edges of the supergranular cells, which primarily consist of strong magnetic flux tubes. The internetwork regions are within the cells where the magnetic field is feeble and weak, and displays barely any significant dynamics. To make this distinction in the observational perspective, we have utilized line of sight (LOS) magnetogram, and the magnetic field inclination data provided by HMI. Figure \ref{fig:fig1}(c) displays the LOS photospheric magnetic field, while the panel (d) shows the inclination of the magnetic field. The magnetic field strength of the ROI is very weak (i.e., magnitude is less than 24 G), and inclination angle lies in the range of 80\textdegree\ to 110\textdegree. Therefore, the region along the slit is essentially the field free region (i.e. internetwork region). We presented the line-of-sight magnetograms and magnetic field inclination maps for visual clarity and to make the distinction of the chosen quiet-Sun regions. Magnetic field information is not utilized in the time-series analysis as these photospheric magnetic field measurements are just contextual  as utilized in the present study. %as the selected region is essentially a very weak field one.
We have chosen two different regions to study the oscillations (i.e., periodic variations in the Doppler velocity time-series) at multiple locations in the solar atmosphere. These locations are analyzed and the physical scenario of the wave propagation is explored. One region spans approximately from y$\approx$0\arcsec to y$\approx$-34\arcsec, and the other region spans approximately from y$\approx$41\arcsec to y$\approx$50\arcsec. To improve the ratio of the signal to the noise in the observed line profiles, we implemented average binning using a $2\times 2$ configuration, involving two bins in time and two bins in $Y-$space. Between the blue marked asterisk as shown on the SJI filter (see, Fig \ref{fig:fig1}), the selected region is located. In addition in the lower region, there is an enhancement of the magnetic field from y$\approx$-24\arcsec to y$\approx$-21\arcsec, which can be seen in the HMI LOS magnetic field. In this region, magnetic field reaches up to 100 G which is quite higher than the surrounding LOS magnetic field. In the appendix \ref{sec:appendix A}, we elaborate the distinct and most peculiar behavior of this magnetically enhanced region vis-\'a-vis associated brightened region in terms of oscillatory power, dominant period, and associated most likely scenario of the wave propagation.

We have used spectral lines Fe I 2799.972 \AA\ formed at 680 km, Mn I 2801.907 \AA\ formed at 830 km \citep{2013ApJ...778..143P}. The formation temperature range of Mg II h 2803.52 \AA\ is log T/K 3.7 - 4.2, while Si IV forms at a higher temperature of log T/K = 4.9. The temperature coverage is given by \cite{2014SoPh..289.2733D}. Fe I 2799.972 \AA\ and Mn I 2801.907 \AA\ are the absorption lines formed in the photosphere \citep[see,][]{2013ApJ...778..143P}. We estimated LOS Doppler velocity at two different heights by fitting both the lines with an inverse single Gaussian function. Mg II h 2803.52 \AA\ is an optically thick line and spans a wide range of altitudes range from the upper photosphere to the upper chromosphere. The line core typically exhibits a central reversal \citep{2013ApJ...772...89L,2013ApJ...772...90L, 2013ApJ...778..143P}. According to \cite{2013ApJ...772...90L}, the Mg II line typically exhibits a double peak in the quiet-Sun, but it can also occasionally display a property of single or multiple peaks. We fitted the Mg II h line using the double Gaussian function as proposed by \cite{2015ApJ...811..127S} and further used by \cite{2019ApJ...874...56R}.

The line core of Mg II h 2803.52 \AA\ is utilized to determine the LOS Doppler velocity. The Si IV 1393.755 \AA\ line is formed in the transition region and is an emission line. A single Gaussian function is employed for fitting the line profile and estimating the LOS Doppler velocity. To accurately estimate Doppler velocities, it is crucial to determine the rest wavelengths. The rest wavelengths of each line is calculated using the shifts of Ni I 2799.474 \AA\ and S I 1392 \AA\ observed in the extreme quiet region{\footnote{\url{https://pyoung.org/quick_guides/iris_auto_fit.html}}}.

\begin{figure*}
\plotone{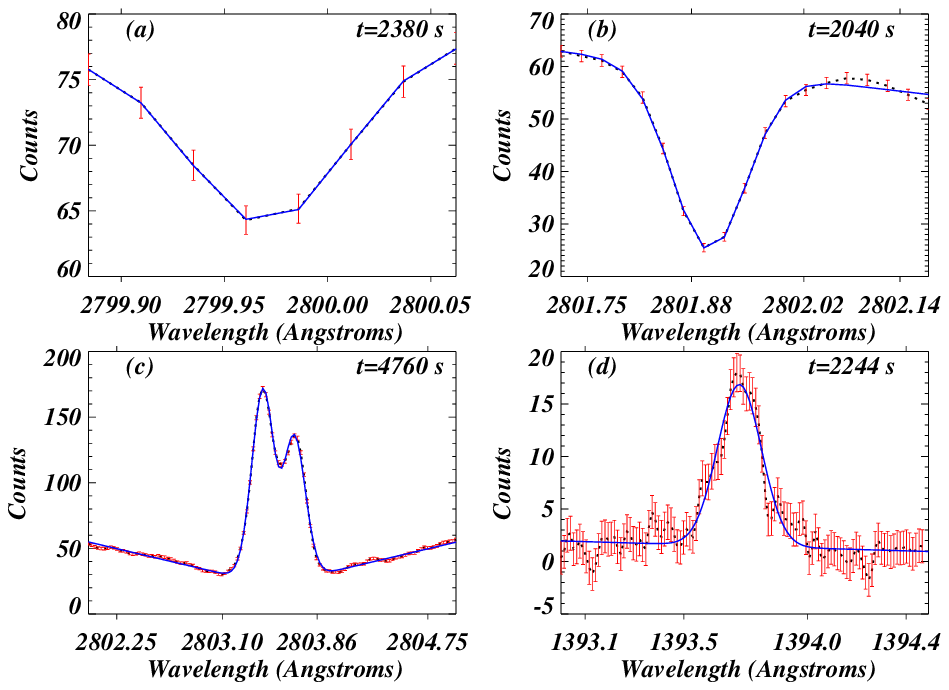}
    \caption{ Spectral fitting of (a) Fe I 2799.972 \AA, (b) Mn I 2801.907 \AA, (c) Mg II h 2803.52 \AA, and (d) Si IV 1393.755 \AA\ lines. The fitted profiles using single Gaussian or double Gaussian function are shown in blue color, whereas the error bars are drawn in red color. These spectral profiles are derived from the location y = 45$\arcsec$, at a given time in sit-n-stare time-series.}
    \label{fig:figure2}
\end{figure*}

Figure \ref{fig:figure2} shows four spectral lines formed at different heights and over-plotted (single or double) Gaussian fittings (Panels (a)-(d)). The black dotted lines are the spectral line profiles, while the overlaid blue color curves represent the corresponding fitted line profiles. 
The error bars are represented by the red color.
%Displayed in red color represents the error bars. 
The observed spectral line profile and the inverse single Gaussian fitting for Fe I 2799.972 \AA\ are presented in Figure \ref{fig:figure2}(a), while Figure \ref{fig:figure2}(b) displays the spectral line profile and the inverse single Gaussian fitting for Mn I 2801.907 \AA. In Figure \ref{fig:figure2}, (c) displays the spectral line profile along with the double Gaussian fitting for Mg II h 2803.52 \AA, while (d) shows the spectral line profile and single Gaussian fitting for Si IV 1393.755 \AA. These example spectral profiles are derived from location y = 45\arcsec at a given time in sit-n-stare time-series. 

There are 145 time bins corresponding to every 'y' location, each containing a spectral line profile. At each of these time bins, we have determined the Doppler velocity signal by fitting the spectral line profile for a particular y-location, and thereby estimating the shift in line centroid w.r.t. the rest wavelength. Some spectra were not available at certain time bins, so we took note of those missing velocity data for each spectral line and they are substituted with the interpolated values. Thus, a time-series of Doppler velocity for each selected $y$ location is generated. The signal-to-noise ratio for the photospheric lines, Fe I and Mn I, was good, as spectra at each time-bin is containing sufficient information. Consequently, there was no need for an interpolation to calculate the Doppler velocity time series. However, for the chromospheric line Mg II h, although the signal-to-noise ratio was generally good, there were instances where we observed triple or single peaks in the spectral line profiles. Therefore, we took note of those spectra and performed interpolation to determine the Doppler shift at those specific bins in order to compute the time series. Conversely, for the TR line Si IV, the signal-to-noise ratio was not reasonable, and there were certain time bins where we were unable to distinguish the spectral peaks from the noise background. Consequently, we documented those time-bins and interpolated the Doppler shift at those points. We repeated this process for 100 different 'y' locations and produced Doppler velocity time-series at each 'y' location at different altitudes, pursuing the analysis of photosphere, chromosphere, and TR spectral lines. Figure \ref{fig:figure3} shows a representative example of time-series for photospheric spectral lines Fe I 2799.972 \AA\ and Mn I 2801.907 \AA\ (upper panel). Similarly in Fig \ref{fig:figure4}, panel (a) shows a representative example of time-series for chromospheric  spectral line Mg II h 2803.52 \AA\ and panel (d) shows a representative example of time-series for TR spectral lines Si IV 1393.755 \AA. These time-series are derived at location y = -2\arcsec.

\subsection{Wavelet analysis of velocity time-series}
\label{sec:wavelet}

\begin{figure*}
  	\mbox{
   \centering
   	\includegraphics[width=.95\linewidth]{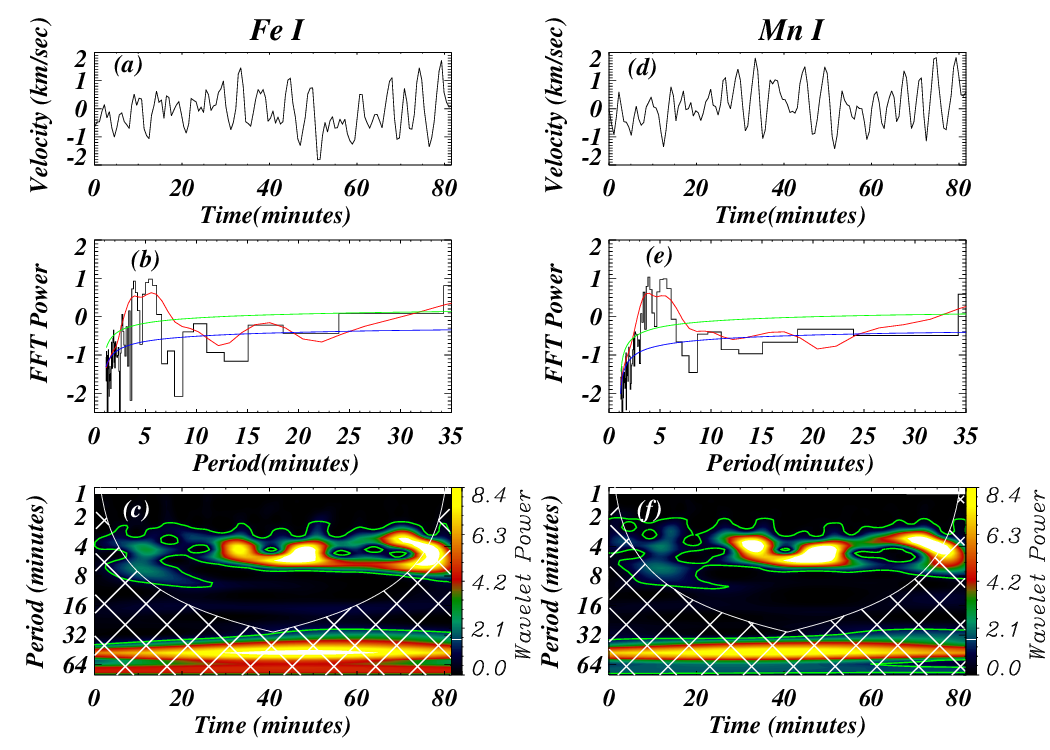} 
   	}

 \caption{Panel (a) shows the time-series of Doppler velocity for Fe I line, while panel (b) shows the time-series of Doppler velocity for Mn I line. Both are obtained from the position y = -2\arcsec. In panels (b) and (e), the FFT power is depicted by the black curve while the red curve corresponds to the global wavelet spectrum. The blue curve represents the power law noise model fit. The 95 \% local significance level is shown in green color. Panels (c) and (f) present the wavelet power spectra for Fe I and Mn I, respectively, with green line contours overlaid representing the 95 \% local significance level.}
   \label{fig:figure3}
\end{figure*}

We used wavelet tool in order to compute the significant periods present in velocity time-series derived from multiple chosen locations at various altitudes (i.e., photosphere, chromosphere, and TR). The wavelet is a well established tool and extensively utilized for detecting periodicities in the time series data \citep{1998BAMS...79...61T}. Wavelet analysis enables the identification of oscillatory powers across both in the frequency and time. Both the Fourier and wavelet analyses account for the detection of oscillatory power, but wavelet analysis differs as its basis functions are localized in both frequency and time, whereas Fourier analysis only localizes basis functions in frequency. For each selected position along the slit (Fig \ref{fig:fig1}, top-right panel), we have a LOS velocity time-series measured at different heights (i.e., in different spectral lines) and we have computed wavelet power spectrum for those different time-series using Morlet wavelet transformation \citep{1998BAMS...79...61T}. For example, the lower left panel of Figure \ref{fig:figure3} displays the wavelet transformation of a representative time series associated with the Fe I spectral line, while the lower right panel exhibits the wavelet transformation of a time series associated with the Mn I spectral line. On the similar basis, Figure \ref{fig:figure4} demonstrate the wavelet tranformation of the Mg II h (lower-left panel) and Si IV (lower-right panel). All these representative time-series are obtained from the position y= -2\arcsec. The cross-hatched area in white is referred to as the cone of influence. Power values falling within this area are deemed insignificant due to the edge effects \citep{1998BAMS...79...61T}. In order to ensure that the obtained power values are reliable, it is necessary to generate significance level contours. The white noise and red noise models are commonly employed as theoretical models to ascertain the levels of significance \citep{1998BAMS...79...61T}. Nonetheless, it should be recognized that these models may occasionally fail to precisely detect the significant power \citep{Auch_re_2016,2020A&A...634A..63K}. Hence, we utilized the power law noise model to determine the significance level instead of relying on the traditional noise models. We have used this noise model in our previous work also \citep[see,][]{2022MNRAS.517..458S}. \cite{Auch_re_2016} introduced the power law noise model, and several authors \citep[e.g.,][]{2017SoPh..292..165T,2020A&A...634A..63K} have used this model in their analysis. The equation for power law noise is expressed as

 \begin{equation} \label{Eq1}
  \sigma(\nu) = A\nu^s + C
\end{equation}

We have utilized this power law model to fit the Fast Fourier Transformation (FFT) of the time-series (see, eq \ref{Eq1}) and we have taken the fitted curve as a background theoretical model and calculated the 95 \% significance level \citep[e.g.,][]{2002SoPh..209..265S,Auch_re_2016,2020A&A...634A..63K}. The over-plotted green contour on the wavelet power spectrum in Fig. \ref{fig:figure3} are the 95\% significance level. 

To compute the periodicity of a time-series we did not take global period instead we have applied certain conditions to the wavelet power of a time-series. These conditions are as follows: (i) Ensuring the power fall within the 95\% local significant contour, (ii) Ensuring the power lying outside the Cone of Influence (COI) in order to mitigate any potential edge effects, (iii) Only those regions are considered where power is significant for three life cycles. After applying these conditions, we extracted the significant powers present in the time series and recorded all associated significant periods corresponding to these significant powers. This exercise has been performed on the time-series derived from 100 chosen locations. Henceforth, we do not bother about the identification of the chosen positions in the quiet-Sun, but assimilated the significant periods and corresponding powers present in all the time-series.

\subsection{Cross correlation between two Doppler velocity time-series originating from different heights}
\label{sec:cross_power_TR}

We use cross-wavelet analysis (e.g., cross-power wavelet, wavelet coherence, and phase difference) to determine the correlations between upper atmospheric oscillations (chromosphere and TR) and lower atmospheric oscillations (photosphere) present in the velocity time-series at a particular chosen location. This procedure is repeated for all the chosen locations in the quiet-Sun (Fig. \ref{fig:fig1}). These estimations are done for two velocity signals from the same pixel location but at different heights. We have performed the cross-wavelet between the spectral line with the minimum height of formation and the other lines, i.e, Fe I - Mn I, Fe I - Mg II h, and Fe I - Si IV. Cross-power analysis identifies significant and shared power in the time-frequency domain between two signals. Wavelet coherence analysis reveals regions of signal coherence, though it does not necessarily indicate common power. A wavelet coherence value of zero indicates no correlation, while a value of one denotes the highest correlation \citep[e.g.,][]{1998BAMS...79...61T,Bloomfield_2004}. The real and imaginary parts of cross-wavelet power help us to calculate phase differences between two signals \citep[e.g.,][]{1998BAMS...79...61T,Bloomfield_2004,2018MNRAS.479.5512K}. In the solar atmosphere, phase difference analysis provides insights into wave propagation. The phase difference reflects the phase lag between the velocity time-series originating from the different heights. Positive phase lags indicate upward wave propagation, whereas negative phase lags denote downward wave propagation between the two heights \citep[e.g.,][]{2016ApJ...819L..23W,2018MNRAS.479.5512K,2020A&A...634A..63K,2020A&A...642A..52A}. In Fig. \ref{fig:figure6}, we presented the result of such analyses for the representative location at y=-2\arcsec. In order to ensure greater reliability of the results, we used certain conditions while selecting the valid phase differences, which are as follows: (i) Only those regions are taken where cross-power is significant for three life cycles, (ii) Ensuring the cross-power falls within the 95\% local significant contour, (iii) Cross-power falling inside the COI is excluded to avoid cross-edge effects, (iv) Wavelet coherence threshold of 0.6 is taken, i.e., only those powers for which wavelet coherence exceeds 0.6 are considered. Section \ref{sec:maximum_oscillatory_powers_TR} presented the scientific results in detail.
\\
\\

%%%%%%%%%%%%%%%%%%%%%%%%%%%%%%%%%%%%%%%%%%%%%%%Fig.2%%%%%%%%%%%%%%%%%%%%%%%%%%%%%%%%%%%%%%%%%%%%%%%%%%%%%%%%%%%%%%%%%%%%%%%%%%%%%%%%%%%%%%%%%%%%%%%%%%%%%%%%%%%%%%%%%%%%%%%%%%%%%%%%%%%%%%%%%%%%%%%%%%%%%%%
%%%%%%%%%%%%%%%%%%%%%%%%%%%%%%%%%%%%%%%%%%%%%%%%%%%%%%%%%%%%%%%%%%%%%%%%%%%%%%%%%%%%%%%%%%%%%%%%%%%%%%%%%%%%%%%%%%%%%%%%%%%%%%%%%%%%%%%%%%%%%%%%%%%%%%%%%%%%%%%%%%%%%%%%%%%%%%%%%%%%%%%%%%%%%%%%%%%%%%%%%%
 
%%%%%%%%%%%%%%%%%%%%%%%%%%%%%%%%%%%%%% Fig 3 %%%%%%%%%%%%%%%%%%%%%%%%%%%%%%%%%%%%%%%%%%%%%%%%%%%%%%%%%%%%%%%%%%%%%%%%%%%%%%%%%%%%%%%%%%%%%%%%%%%%%%%%%%%%%%%%%%%%%%%%%%%%%%%%%%%%%%%%%%%%%%%%%%%%%%%%%%%%%

%%%%%%%%%%%%%%%%%%%%%%%%%%%%%%%%%%%%%%%%%%%%%%%%%%%%%%%%%%%%%%%%%%%%%%%%%%%%%%%%%%%%%%%%%%%%%%%%%%%%%%%%%%%%%%%%%%%%%%%%%%%%%%%%%%%%%%%%%%%%%%%%%%%%%%%%%%%%%%%%%%%%%%%%%%%%%%%%%%%%%%%%%%%%%%%%%%%%%%%%%%%%

\section{Results}

\label{sec:maximum_oscillatory_powers_TR}

\begin{figure*}
  	\mbox{
   \centering
   	\includegraphics[width=.95\linewidth]{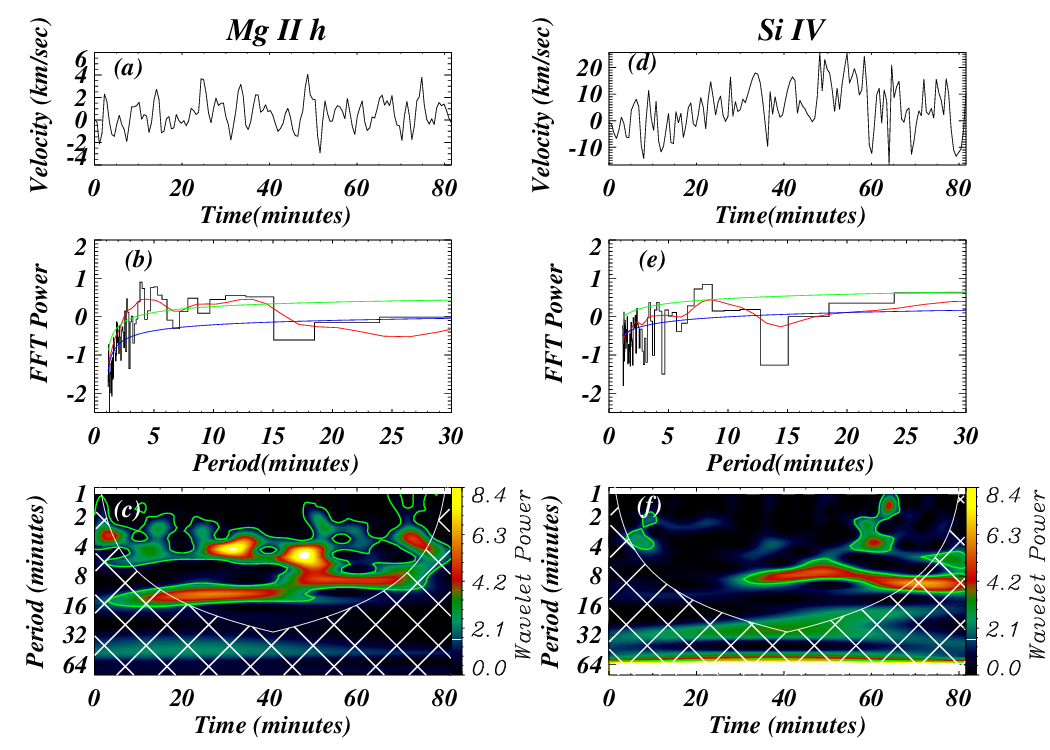} 
   	}
  \caption{The left panel shows the Doppler velocity, model fit, and wavelet power spectrum for the Mg II h line, while the right panel displays the same features for the Si IV line. The descriptions for panels (a)-(f) are analogous to those provided in Figure \ref{fig:figure3}.}
  \label{fig:figure4}
\end{figure*}

\begin{figure*}
  	\mbox{
   \centering
   	\includegraphics[width=.95\linewidth]{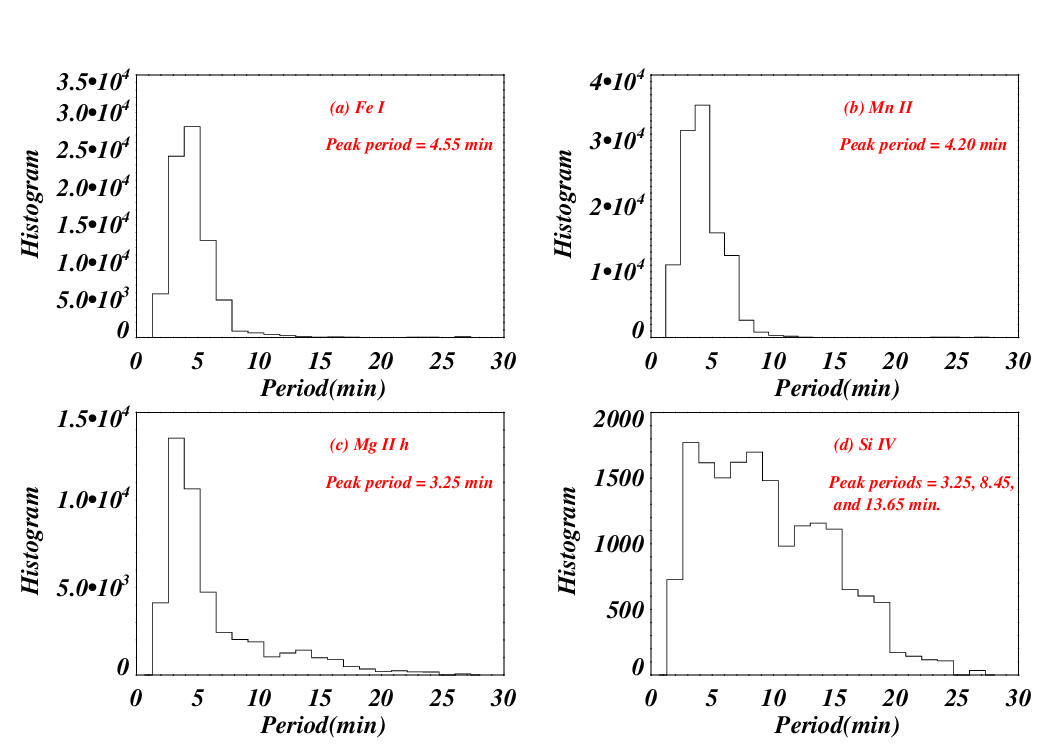} 
   	}
  \caption{Panel (a) displays the period distribution for the Fe I line, while panel (b) shows the period distribution for Mn II. In panel (c), the period distribution for Mg II h is presented, and panel (d) shows the period distribution for Si IV.}
  \label{fig:figure5}
\end{figure*}

Here, we have presented the results from one representative example from a single location (y = -2\arcsec) in Fig.~\ref{fig:figure3} and Fig.~\ref{fig:figure4}. In Fig. \ref{fig:figure3}(c), there are two power patches lie within the 95\% local significance level. One is around 2-6 min and another is around 32-64 min. The second patch lies within the COI, hence this power is not included in further analysis (see more details in section \ref{sec:wavelet}). Similarly, for Mn Doppler velocity time-series, the wavelet power highlights a significance within 1.5-6 minute period range. In Fig. \ref{fig:figure4}, for the Mg II h time series, the significant power is observed in 2-4 min period range (panel c), while for the Si IV spectral line, it ranges between 2-8 min. Similar to this representative case, we have performed the wavelet analysis of velocity time-series derived from all selected locations between $\approx$-34\arcsec to y$\approx$0\arcsec and y$\approx$41\arcsec to y$\approx$50\arcsec~for each spectral line and derived our results based on the statistical analysis. We have applied certain conditions listed in section \ref{sec:wavelet} to the power spectral analysis (i.e., wavelet) of each time-series and and stored all significant periods corresponding to the significant powers. Subsequently, we plotted a 1-D histogram of stored significant periods for different lines as shown in Fig. \ref{fig:figure5}. The period distributions for Fe I and Mn I spectral lines are displayed in Fig. \ref{fig:figure5} (a) and (b), respectively. The peak period for Fe I spectral line is 4.55 min, while the peak period for Mn I spectral line is 4.20 min. Figure \ref{fig:figure5} (c) shows the distribution of periods in the Mg II h spectral line and it is clearly visible in the histogram that 3 min period is dominant in the chromospheric region. The 3-minute period oscillations observed in the chromosphere could potentially be linked to the underlying photospheric oscillations. To determine the origin of these chromospheric oscillations, we conducted a cross-correlation analysis between the photospheric and chromospheric oscillations, as outlined in the sub-section \ref{sec:cross_power_TR}. Figure \ref{fig:figure5} (d) shows the distribution of periods in the Si IV spectral line and these periods are distributed within the entire range between 2-30 min. In our previous paper, we have also found a similar kind of distribution in the transition region (TR), i.e., significant periods are distributed in 1-20 min range of periods \citep[see,][]{2022MNRAS.517..458S}. In the present case, we have found dominance of shorter periods along with the presence of longer periods in the period distribution as evident in Si IV TR line. In the transition region, the period peaks at 3.25, 8.45, and 13.65 min. As discussed in the next section, shorter periods ($<$3.0 min) may be correlated with the photospheric oscillations, while the longer periods may be generated {\it in-situ} in TR.

%%%%%%%%%%%%%%%%%%%%%%%%%%%%%%%%%%%%%%%%%%%%%%%%%%%%%%%%%%%%%%%%%%%%%%%%%%%%%%%%%%%%%%%%%%%%%%%%%%%%%%%%%%%%%%%%%%%%%%%%%%%%%%%%%%%%%%%%%%%%%%Fig4%%%%%%%%%%%%%%%%%%%%%%%%%%%%%%%%%%%%%%%%%%%%%%%%%%

%%%%%%%%%%%%%%%%%%%%%%%%%%%%%%%%%%%%%%%%%%%%%%%%%%%%%%%%%%%%%%%%%%%%%%%%%%%%%%%%%%%%%%%%%%%%%%%%%%%%%%%%%%%%%%%%%%%%%%%%%%%%%%%%%%%%%%%%%%%%%%%%%%%Fig5%%%%%%%%%%%%%%%%%%%%%%%%%%%%%%%%%%%%%%%%%%%%%%%%%%

%%%%%%%%%%%%%%%%%%%%%%%%%%%%%%%%%%%%%%%%%%%%%%%%%%%%%%%%%%%%%%%%%%%%%%%%%%%%%%%%%%%%%%%%%%%%%%%%%%%%%%%%%%%%%%%%%%%%%%%%%%%%%%%%%%%%%%%%%%%%%%%%%%%%%% Fig 6%%%%%%%%%%%%%%%%%%%%%%%%%%%%%%%%%%%%%%%%%%%
\begin{figure*}
  	\mbox{
   \centering
   	\includegraphics[width=.95\linewidth]{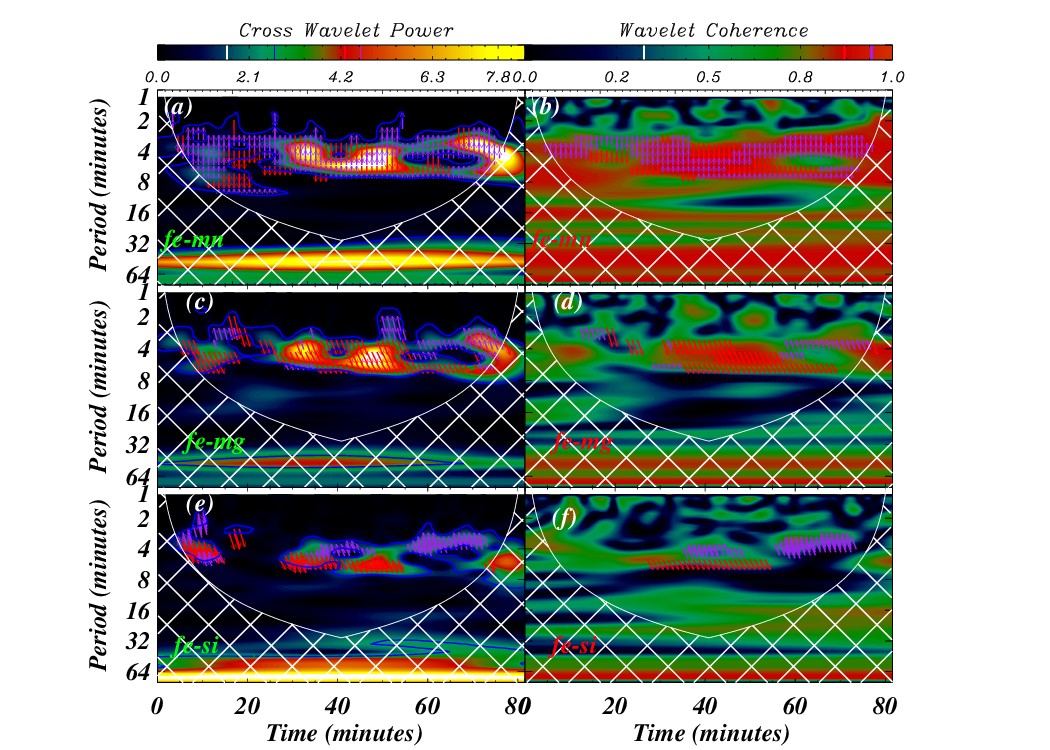} 
   	}
 \caption{Panels (a) and (b) show the cross-power wavelet and wavelet coherence between Fe I and Mn I velocity signals, while panels (c) and (d) display the same features for the Fe I - Mg II pair. Panels (e) and (f) display the same for Fe I and Si IV panels. The left panels feature an over-plotted blue line contours representing 95 \% local significance level. Power values falling within this contour are considered significant. Arrows overlaid on the cross-power map indicate the phase differences between the two velocity signals derived from the representative chosen location y=-2\arcsec. The purple arrows represent a positive phase lag and red arrows depict a negative phase lag.}
  \label{fig:figure6}
\end{figure*}

%%%%%%%%%%%%%%%%%%%%%%%%%%%%%%%%%%%%%%%%%%%%%%%%%%%%%%%%%%%%%%%%%%%%%%%%%%%%%%%%%%%%%%%%%%%%%%%%%%%%%%%%%%%%%%%%%%%%%%%%%%%%%%%%%%%%%%%%%%%%%%%%%%%%%%%%%%%%%%%%%%%%%%%%%%%%%%%%%%%%%%%%%%%%%%%%%%%%%%%%%%

\begin{figure*}
  %% \centering
   	
\includegraphics[width=.9\linewidth]{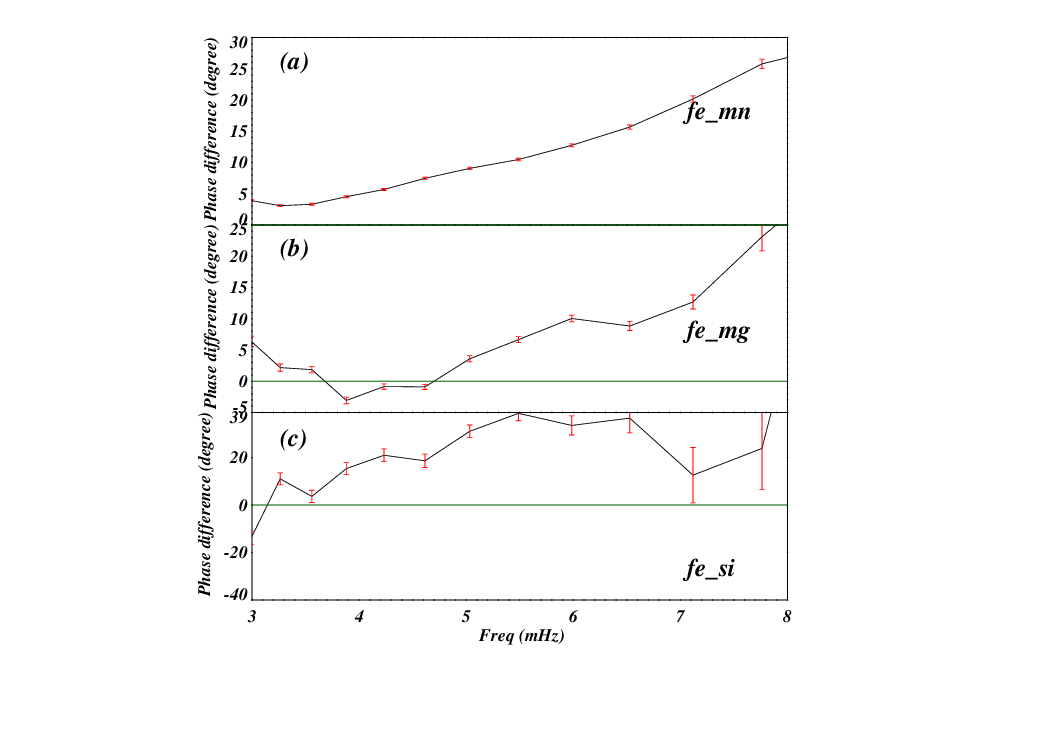}
\caption{Panel (a) displays the phase difference with frequency variation for the Fe I - Mn I pair, while panel (b) shows the phase difference with frequency variation for the Fe I - Mg II h pair. Panel (c) describes the phase difference with frequency variation for the Fe I - Si IV pair.}
\label{fig:figure7}
\end{figure*}

The statistically significant oscillatory periods present at different heights may be associated with the linear propagation of magnetoacoustic waves. To confirm this physical scanario we perform the cross correlation analysis as discussed in sub-section \ref{sec:cross_power_TR}. 
In Fig. \ref{fig:figure6}, the violet arrows represent the positive phase difference and hence indicate the upward propagation of the waves. The red arrows represent the negative phase difference and hence indicate the downward propagation of the waves. For pair Fe-Mn, the significant cross-power lies in the range of 2-6 min periodicities as this power lies within 95\% significance contours. For Fe I-Mg II h pair, significant cross-power lies in the range of 2-5 min, while for Fe-Si pair significant cross-power lies in the range of 2-5 min. We have applied certain conditions (see section \ref{sec:cross_power_TR}) and sorted the reliable phase differences for different line pairs as mentioned above. For visualization, we have plotted arrows of reliable phase differences on the wavelet coherence map after imposing those conditions in panels (b), (d), and (f) of Fig. \ref{fig:figure6}. In the case of Fe-Mn pair (see, panel (a)), a significant cross-power exists within the period range of 8-12 in the time domain of 10-30 minutes. However, panel (b) lacks phase difference arrows related to this cross-power, indicating the absence of reliable phase differences for this specific cross-power. Consequently, we have not considered the phase differences corresponding to this cross-power in our further analysis. Similarly, for other pairs like Fe-Mg (panel (d)) and Fe-Si (panel (f)), we have also identified non-reliable phase differences and removed them from further analysis. This example is related to the representative chosen location at y=-2\arcsec. These conditions were also applied across all the chosen locations, and we have extracted all reliable phase differences between different line pairs at each location in the time-frequency domain. Subsequently, we calculated the mean and standard errors of the phase differences at each period for different line pairs. The mean phase lag represents the phase difference between two signals originating from different heights over a period, and its standard error represents the phase error. We then plotted the phase difference versus period for each pair of lines and plots are shown in Fig. \ref{fig:figure7}. A black color curve shows variation in phase difference, whereas a red vertical line shows an error in phase difference. Panel (a) displays the phase difference versus period plot for the Fe I - Mn I pair, while Panel (b) corresponds to the Fe I - Mg II h pair, and Panel (c) corresponds to the Fe I - Si IV pair. By analyzing the phase difference variation at different frequencies between the two distinct heights, we have estimated the cut-off frequency at various heights.

We defined the cut-off frequency as the point at which the phase difference becomes greater than zero, and where there is a subsequent increase in the phase difference with frequency. For the pair Fe I - Mn I (see, Fig. \ref{fig:figure6} (a)), the positive value of phase difference is obtained, so there is no cut-off frequency found at this height difference, and all waves within the frequency range 3-8 mHz (or period range 2-6 min, thus including 3 min and 5 min) can propagate from the lower photosphere to the upper photosphere. For the pair Fe I - Mg II h (see, Fig. \ref{fig:figure6} (b)), the phase difference starts increasing at frequency 4.7 mHz (or 3.54 min period), hence the cut-off frequency for this height difference is 4.7 mHz (or cut-off period 3.54 min). Hence, waves with a frequency greater than 4.7 mHz (or with a period less than 3.54 min) can propagate to the chromospheric height . Thus only the wave with 3-min period is propagating from the photosphere to the chromosphere, while the wave with 5-min period is propagating in the downward direction, from the chromosphere. For the pair Fe I - Si IV (see, Fig. \ref{fig:figure6} (c)), the phase difference starts increasing at 3.14 mHz (or 5.3 min), hence  cut-off frequency for this height difference is 3.14 mHz (or cut-off period 5.3 min). Hence, waves with a frequency greater than 3.14 mHz (or with a period less than 5.3 min) can propagate to the TR height. That means the waves with shorter periods (3 min) may propagate from chromosphere to TR region hence correlated with the photospheric oscillations, while the longer periods waves are generated {\it in situ} between chromosphere and the transition height, however, certainly not by the photospheric oscillations. These observed cut-off frequencies as representing the minimum values that indicate the actual cut-off periods between the respective heights being measured.

\citet{2018MNRAS.479.5512K} studied oscillations in the internetwork region and reported almost similar behaviour of the oscillations. They detected the presence of 3-min oscillations occurring in both the chromosphere and the Transition Region (TR) within the internetwork regions. They concluded that these oscillations were directly influenced by the photospheric oscillations. Furthermore, they reported the existence of higher wave periods ranging from 2.5 to 6 min specifically in the TR. Importantly, they observed that the higher periods observed in the chromosphere, particularly within the 5-min oscillation range, served as a source for the corresponding higher periods detected in the TR. These findings suggest a relationship between the chromospheric and TR oscillations, with the chromosphere acting as a potential driver for the low-frequency (or high period) oscillations observed in the TR.
Our analysis clearly demonstrates the presence of a height dependence of cut-off frequency (or cut-off period). This observation also follows previous studies conducted by \citet{2016ApJ...819L..23W} and \citet{2019A&A...627A.169F}, who have reported similar trend regarding the height dependence of cut-off frequency/period. An example in Fig. \ref{fig:figure7} demonstrates that in Fourier-Period \& Real Time domain, the propagation behaviour of the waves exhibits highly complex physical scenario. At a given representative location (y=-2\arcsec), this analysis of phase difference (+ve and/or -ve) reveal the presence of both upward and downward propagating waves associated with different Fourier periods, else also the varying nature of propagation on the same Fourier period in different epoch/segment of time. This embarks that at a given particular location, the complex scenario of the wave propagation is present. Somewhat, such trends do appear at all the analyzed spatial locations.  Some upward propagating waves may follow the average cut-off frequency threshold, while others may not obey it depending upon exclusive plasma and magnetic field conditions present there. There may be a multitude of physical conditions that may yield such a complex behaviour of the wave propagation. It may be related to the magnetic field \citep[e.g.,][]{2002ApJ...564..508R}, or non-adiabaticity of the waves \citep[e.g.,][]{2023ApJ...949...99W}, or maybe due to the wave reflections from strong gradients at the TR \citep[e.g.,][]{2021RSPTA.37900170F}. However, the present observational base-line can not comprehend them directly, except the study of cut-off and wave propagation scenario. In Appendix \ref{sec:appendix C}, we have shown diverse wave propagation behavior at some locations. However, there average behaviour over the chosen large-scale quiet-Sun ({\it viz.} aligning with the locations chosen over the IRIS slit) follows the classical physical condition of the cut-off permeated medium where certain selective frequencies can only propagate up (Fig \ref{fig:figure7}).

%%%%%%%%%%%%%%%%%%%%%%%%%%%%%%%%%%%%%%%%%%%%%%%%%%%%%%%%%%%%%%%%%%%%%%%%%%%%%%%%%%%%%%%%%%%%%%%%%%%%%%%%%%%%%%%%%%%%%%%%%%% Fig 7%%%%%%%%%%%%%%%%%%%%%%%%%%%%%%%%%%%%%%%%%%%%%%%%%%%%%%%%%%%%%%%%%%%%%%%%%%
%%%%%%%%%%%%%%%%%%%%%%%%%%%%%%%%%%%%%%%%%%%%%%%%%%%%%%%%%%%%%%%%%%%%%%%%%%%%%%%%%%%%%%%%%%%%%%%%%%%%%%%%%%%%%%%%%%%%%%%%%%%%%%%%%%%%%%%%%%%%%%%%%%%%%%%%%%%%%%%%%%%%%%%%%%%%%%%%%%%%%%%%%%%%%%%%%%%%%%%%%%

\section{Discussion and conclusions}
\label{sec:discussion}

We have used data from IRIS instrument to examine wave propagation in the solar atmosphere, by analyzing different spectral lines formed at various heights/temperatures. The power spectral analysis of velocity time-series data from multiple inter-network regions provided insights into average dominant periods across the photosphere to the transition region. The average behavior of the QS internetwork is obviously a classical cut-off period permeated atmosphere where 5 min magnetoacoustic waves dominate from lower to the upper atmosphere, while 3 min wave power is dominant at chromosphere and TR. The higher period oscillations are also presented in TR. In order to find the correlation of these TR oscillations with the photospheric oscillations we have performed cross wavelet analysis (i.e., cross-wavelet analysis, coherence, and phase difference analysis). We observed that shorter period oscillations are likely due to photospheric oscillations, while longer period oscillations are generated \textit{in-situ} on the basis of cut-off frequency estimation. We determined the cut-off frequencies between different atmospheric regions, finding no cut-off frequency in the upper photosphere, a cut-off frequency of 4.7 mHz (or cut-off period of 3.54 min) between the chromosphere and lower photosphere, and a cut-off frequency of 3.14 mHz (or cut-off period of 5.3 min) between the transition region and lower photosphere. \cite{2009ApJ...696.1448W} also studied Doppler and intensity oscillations in the transition-region line (He ii) and five coronal lines (Fe x, Fe xii, Fe xiii, Fe xiv, and Fe xv). They first estimated the evolution of Doppler shift and intensity averaged over 11 pixels, then performed the analysis. \cite{2010ApJ...713..573M} also averaged Doppler shift and intensity over 16 pixels, then performed the analysis. They commented on the physical nature of these oscillations. However, in our analysis, we did not use averaged values; instead, we performed the wavelet analysis tool for each chosen location (i.e., pixel). We found that the wave propagation behavior varies across different locations and exhibits a very complex scenario with upward and downward propagating waves at different times (more details in Appendix \ref{sec:appendix C}). Therefore, we suggest that analyzing single or few spatial locations will not be sufficient to comprehend the true physical scenario of the magnetoacoustic wave propagation in the solar atmosphere.

\cite{10.1112/plms/s2-7.1.122} has firstly derived the formula for cut-off frequency. He assumed an isothermal atmosphere, which led him to regard the cutoff frequency as a global quantity. For an atmosphere like the solar atmosphere, a more realistic assumption must be made. The choice of wave equations in several theoretical models led to different cut-off frequencies \citep[e.g.,][]{1984ARA&A..22..593D,1998A&A...337..487S,2006PhRvE..73c6612M}.
Previously, the analysis of power spectra is used to estimate the acoustic cut-off frequency. This approach has produced a wide range of values ranging from 5.3 to 5.7 mHz. However, determining the cutoff frequency has become more difficult as modulated signals beyond this frequency range are observed. Therefore, relying on a sudden drop in power density is no longer an effective method. \cite{2006ApJ...646.1398J} employed bivariate analysis (different from the power spectra approach) using coherence and phase shift to estimate the cut-off frequency, and found a value of 5.1 mHz. This value is in close proximity to the previously reported value of 5.3 mHz.
\cite{2016ApJ...819L..23W} and \cite{2019A&A...627A.169F} have studied cut-off frequency variation with atmospheric height in different solar atmospheric regions using observational data. The study by \cite{2016ApJ...819L..23W} investigated the variation in the cut-off frequency within the quiet-Sun region. They have estimated the phase difference spectra between velocity signals from the lines with the highest formation height and the rest lines.
They have found an increasing trend of the cut-off frequency with the atmospheric height. They have also compared the variation of the cut-off frequency computed using observational data with the cut-off frequency variation computed using the theoretical model. \cite{2019A&A...627A.169F} have also conducted a study on height-dependent variation of the cut-off frequency of slow magnetoacoustic waves in a sunspot umbra. Phase differences and amplification spectra were computed between velocity signals obtained from various pairs of spectral lines formed in the solar atmosphere. The results revealed a cut-off frequency of 5 mHz at the deep photosphere, 6 mHz at the upper photosphere, and 3 mHz at the chromospheric altitude. These findings conclusively demonstrate that the cut-off frequency is height-dependent. \cite{2018MNRAS.479.5512K} have investigated the propagation of acoustic waves in the quiet-Sun. They did not estimate the variation of cut-off frequency with the height. Instead, they have studied the cross-correlation between the oscillation signals at two different heights. They have found oscillations within period regime 1.6 to 4.0 min propagating into the chromosphere from the lower atmosphere, i.e., the photosphere, while oscillations having a period above 4.5 min propagate downward. Therefore they have suggested a 4 min cut-off period at the chromospheric height. As a result, they have also demonstrated indirectly an evidence of cut-off frequency (or cut-off period) in the solar atmosphere.

The importance of the acoustic cut-off frequency lies in its role in establishing the conditions under which waves propagate, and also in identifying regions within the solar atmosphere where waves exhibit strong reflection. The observed cut-off period of 3.5 minutes is in line with the classical theory of acoustic wave propagation, as supported by earlier studies \citep{1932hydr.book.....L,1993A&A...273..671F}. The selected region for studying wave propagation is essentially magnetic field free, as indicated by the distribution of line-of-sight photospheric magnetic fields along the IRIS slit. Therefore, we conjecture the fundamentally acoustic nature of the waves in these regions. Previously, various studies has been conducted to investigate the propagation of acoustic waves through both observational methods and numerical simulations \citep{1979ApJ...231..570L,1997ApJ...481..500C, 2001ApJ...554..424J,2016ApJ...827...37M}. The 3-minute oscillations (i.e., 5.5 mHz) exhibit lower power compared to 5-minute oscillations (3.3 mHz) at the solar photosphere, they predominate in the chromosphere, where 5-minute oscillations become evanescent \citep[e.g.,][]{1979ApJ...231..570L,1982ApJ...258..393L,1997ApJ...481..500C,2016ApJ...827...37M}. However, the presence of an inclined or strong magnetic field may leads to the reduction of the cut-off frequency. For instance, Numerical simulations conducted by \cite{2005ApJ...624L..61D} have shown that 5-minute photospheric oscillations can effectively propagate into the corona when they follow an inclined magnetic flux tube. They found that the cut-off frequency is dependent on the inclination of the magnetic field. In non-vertical magnetic fields, the cut-off frequency increases, allowing 5-minute waves to penetrate into the upper atmosphere. \cite{2001A&A...371.1137B} has studied the network oscillations (the region with strong magnetic field), utilizing lines formed at varying temperatures, and has indicated that these oscillations are a result of wave activity. Moreover, the analysis has revealed that the dominant period of the transition region and chromospheric lines falls within the range of 2-4 mHz. It is possible that p-modes act as the driving force behind oscillations observed in the transition region and chromosphere, while the magnetic field appears to have a significant impact on these oscillations.

We have observed longer period oscillations in the chromospheric and TR lines at certain locations, which cannot be attributed to photospheric oscillations (details are given in Appendix \ref{sec:appendix A}). \cite{1993ApJ...414..345L} investigated oscillations in the network region of the quiet Sun using photospheric and chromospheric lines and discovered longer periods (exceeding 5 min) dominating above the chromospheric network. Using phase difference analysis, they established that these chromospheric oscillations do not exhibit coherence with oscillations in the underlying photosphere. As a result, these oscillations could either be confined to the chromosphere or may originate from photospheric phenomena. Additionally, \cite{2002ApJ...567L.165M} have reported long-period waves in the range of 4-15 min, attributing them to magnetoacoustic or magnetoacoustic-gravity modes rather than acoustic origin. Several other authors have also reported long-period oscillations in the chromosphere region \citep{1995MNRAS.274L...1L,2000A&A...357.1093C}.

When waves propagate in the solar atmosphere, they may exhibit repeated nonlinear patterns, such as the sawtooth pattern observed in the time series of intensity and Doppler velocities, typically within the chromospheric and transition regions. \cite{2006ApJ...640.1153C} and \cite{2014ApJ...786..137T} have also previously reported this sawtooth pattern while studying umbral oscillations. \cite{1994chdy.conf...47C} and \cite{1997ApJ...481..500C} suggested that these patterns are a result of shocks. The temporal evolution of line profiles, encompassing both chromospheric and transition region spectral line profiles, was analyzed by \cite{2014ApJ...786..137T}. Within their analysis, they discerned a recurrent pattern wherein the line core initially undergoes a sudden impulsive blueward excursion, accompanied by an enhancement in intensity, followed by a gradual and consistent deceleration towards the red side. The line width experiences a sudden increase as the Doppler shift shifts from redshift to blueshift. The correlated alterations between intensity and blueshift in the transition region lines, coupled with time lags and significant nonlinearities, indicate the potential presence of magnetoacoustic shock waves propagating from the chromosphere to the transition region.
 
In their work, they observed a clear sawtooth shape with a slow evolution of velocity towards positive values, followed by a sudden blueshift, indicating the presence of shock waves. We also observed some large amplitude behavior at certain instantaneous times for the representative time-series of Mg II and Si IV, but we did not identify a recurring sawtooth pattern, therefore not an indication of the presence of shock waves (more details are given in Appendix \ref{sec:appendix B}). We conjecture that the wave is trying to steepen at some locations while propagating upward, however, the local atmospheric conditions smoothed out the distortions in the wave profile and made it remaining in an oscillatory phase overall. 

Investigating the velocity oscillations plays some role in comprehending the energy equilibrium of the chromosphere and the magnetic configuration of the outer atmosphere. In general, these studies involve the quantification of energy flow in waves that propagate at various heights of the solar atmosphere. Consequently, it is of paramount importance to accurately characterize the cut-off frequency to determine which frequencies can effectively contribute to the energy transport and heating process in the overlying atmosphere. Our statistical study comprises the presence of dominant frequencies in the large patch of QS, their signature of propagation, estimation of cut-off frequency. We also emphasize a diversity in their physical behavior and propagation properties that added a comprehensive view in the existing understanding of the wave propagation. The observed oscillations primarily possess a magnetic nature, which is why we have labeled them as magnetoacoustic waves. Currently, we lack magnetic field measurements across various atmospheric layers, making it challenging to know where these pure acoustic oscillations convert into magnetoacoustic waves. The magnetic fields estimated by SDO/HMI can only provide values for photosphere only. However, we lack information regarding the field strength and inclination in the chromosphere and transition region. Consequently, quantifying wave behavior based on magnetic field measurements across different layers of the solar atmosphere remains an open question.

\section{acknowledgments}

We acknowledge the valuable remarks from referee that improved our manuscript considerably. KS acknowledges UGC for providing grant for her research work. AKS acknowledges the ISRO Project Grant (DS 2B-13012(2)/26/2022-Sec.2) for the support of his research. We acknowledge the use of IRIS spectral data, AIA imaging data, HMI data and the wavelet tool of Torrence and Compo (1998). IRIS is a NASA small explorer mission developed and operated by LMSAL with mission operations executed at NASA Ames Research Center and major contributions to downlink communications funded by ESA and the Norwegian Space Centre. DY is supported by National Natural Science Foundation of China (NSFC,12173012,12111530078, 11803005), the Guangdong Natural Science Funds for Distinguished Young Scholar (2023B1515020049), the Shenzhen Technology Project (GXWD20201230155427003-20200804151658001), and the Shenzhen Key Laboratory Launching Project (No. ZDSYS20210702140800001).

\vspace{5mm}

%;;;;;;;;;;;;;;;;;;;;;;;;;;;;;;;;;;;;;;;;;;;;;;;;;;;;;;;;;;;;;;;;;;;;;;;

%\end{comment}
%%%%%%%%%%%%%%%%%%%%%%%%%%%Appendix%%%%%%%%%%%%%%%%%%%%%%%%%%%%%%%%%%%%%%

%%%%%%%%%%%%%%%%%%%%%%%%%%%%%% APPENDIX%%%%%%%%%%%%%%%%%%%%%%%%%%%%%%%%
\appendix

\section{Power at Strong Magnetic Patch with Bright Point}
\label{sec:appendix A}

\begin{figure*}
  	\mbox{
   \centering
   	\includegraphics[width=.95\linewidth]{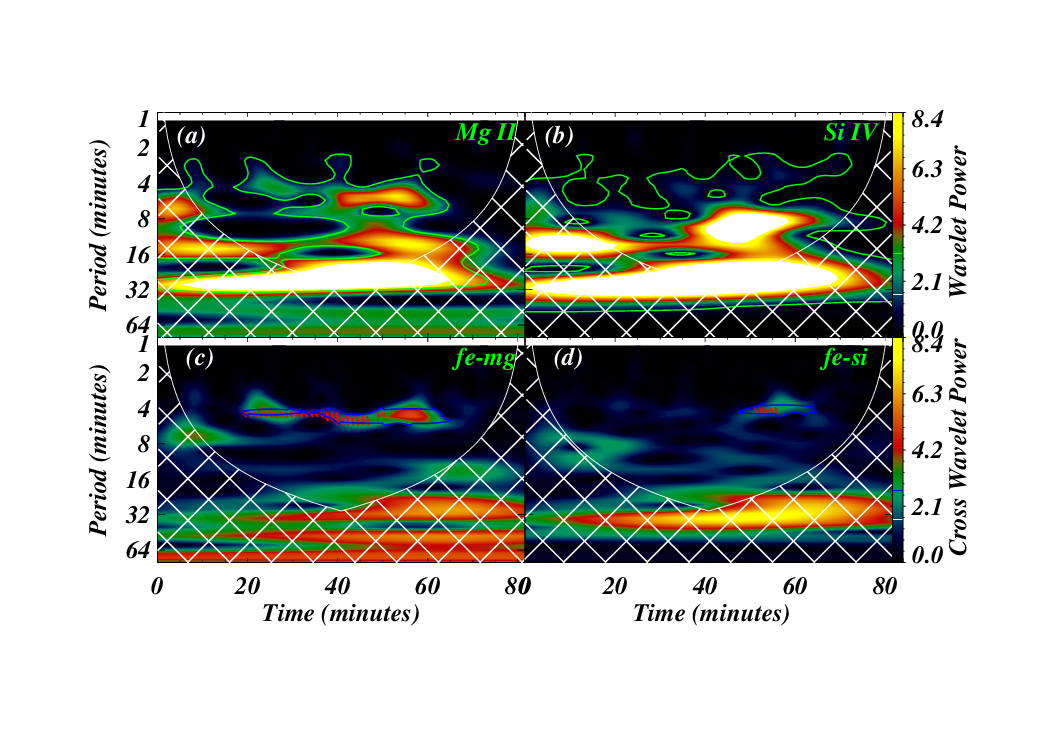} 
   	}
 \caption{Panel (a) displays the wavelet power spectra for the Mg II line, while panel (b) presents the corresponding spectra for the Si IV line. The remaining details are similar to those in Figure \ref{fig:figure3}. In panel (c), the cross-power wavelet between Fe I and Mg II velocity signals is displayed. Panel (d) shows the cross-power wavelet between Fe I and Si IV velocity signals, following the details outlined in Figure \ref{fig:figure6}. It is important to highlight that these analyses focus on velocity time-series derived from strongly magnetized patches associated with bright points, revealing distinct oscillatory behaviors. This location lies at y=-22\arcsec.}
  \label{fig:figure8}
\end{figure*}

We conducted a statistical investigation of wave propagation in the quiet Sun and discovered a prevalence of 3-minute periodicity in the chromosphere region. However, upon examining the HMI magnetogram, we observed an increase in magnetic field strength at certain locations along the slit (highlighted by the red box in Fig. \ref{fig:fig1}). Wavelet analysis revealed the existence of longer periods in the range of 5-30 minutes in the chromospheric and transition region lines. We provided an example to demonstrate the distinct behavior of the oscillations at these locations in Fig. \ref{fig:figure8}. Wavelet power in the Mg II and Si IV lines was primarily dominant at longer periods of 15-32 min in both panels (a) and (b), respectively. Cross-power between velocity signals of Fe I-Mg II and Fe I-Si IV, shown in panels (c) and (d), respectively, demonstrated that longer period oscillations (greater than 5 min) at chromospheric and transition region heights were not correlated with the photospheric acoustic flux. The observed oscillation suggests that it is likely generated in the chromosphere, and the propagating wave could undergo a mode conversion process near the upper chromospheric layer, where the magnetic pressure equals the gas pressure. Additionally, the presence of a strong magnetic field may introduce {\it in situ} origin of wave modes as the atmosphere gets higher, even though the p-modes are primarily acoustic deep down. Numerous studies have reported an increase in wave power at higher altitudes, i.e., chromosphere and transition region, over regions with increased photospheric magnetic field \citep{2001ApJ...548L.237M,2007PASJ...59S.699H}.

%%%%%%%%%%%%%%%%%%%%%%%%%%%%%%%%%%%%%%%%%%%%%%%%%%%%%%%%%%%%%%%%%%%%%%%%%%%%%%%%%%%%%%%%%%%

\section{Non Linear Behavior at different time bin in Mg II h and Si IV spectra}
\label{sec:appendix B}

\begin{figure*}
  	\mbox{
   \centering
   	\includegraphics[width=.95\linewidth]{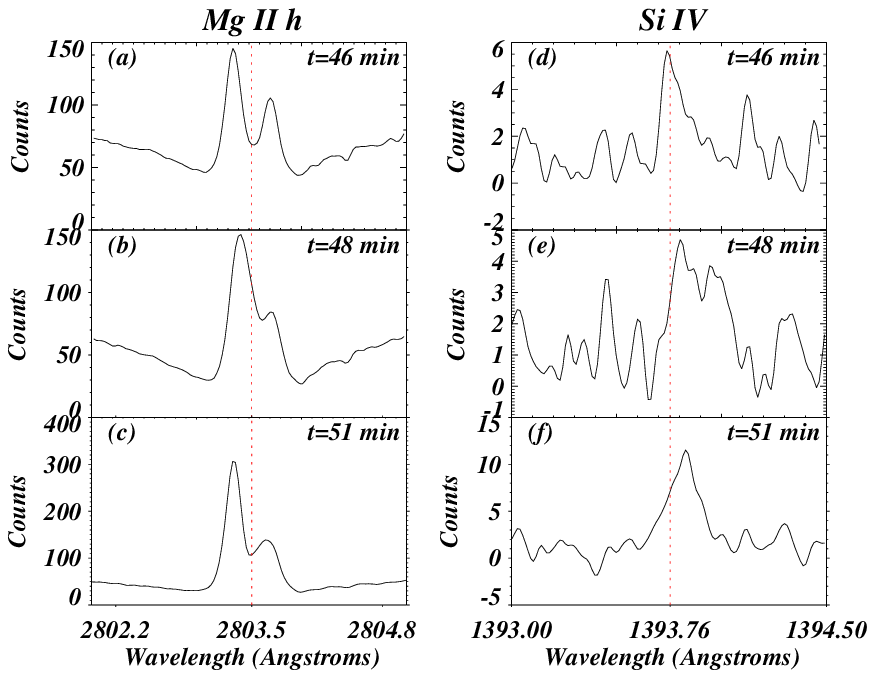} 
   	}
 \caption{In the left panel, the spectra of Mg II h at various time points are displayed, while the right panel show the spectra of Si IV at different time intervals. The rest wavelength of the spectra is indicated by a red vertical dashed line. These spectra are corresponding to different time epoch for the pixel location y=-2\arcsec.}
  \label{fig:figure9}
\end{figure*}

In specific time bins, nonlinear behavior is evident in the representative time-series of the Mg II and Si IV lines, around the 50 minute in Figure \ref{fig:figure4}. It should be noted that these time bins are derived from y=-2\arcsec. The corresponding spectra for these time bins of the Mg II and Si IV lines are shown in Figure \ref{fig:figure9}. In the case of the Mg II h line profile, there is a shift to redshift followed by blueshift, while in case of Si IV profile, there is an evolution of velocity towards positive values with an increase in line width, followed by shift of velocity towards negative value. These are indicative of some nonlinear behavior, but these profiles do not show any sawtooth shape as reported in previous literatures \citep[e.g.,][]{2006ApJ...640.1153C, 2014ApJ...786..137T}. Hence, this non linear behavior is not an indication of the presence of shock waves in the representative time-series. Such distinctive large amplitude velocity oscillation are present at any particular time epoch in the time-series data derives from various locations also. However, they cannot considered as a signature of the shock wave trains. The derived velocity oscillations in time are merely associated with magneto-acoustic wave propagation.

\section{Example of distinct behavior of propagating waves at different locations}
\label{sec:appendix C}

\begin{figure*}
  	\mbox{
   \centering
   	\includegraphics[width=.95\linewidth]{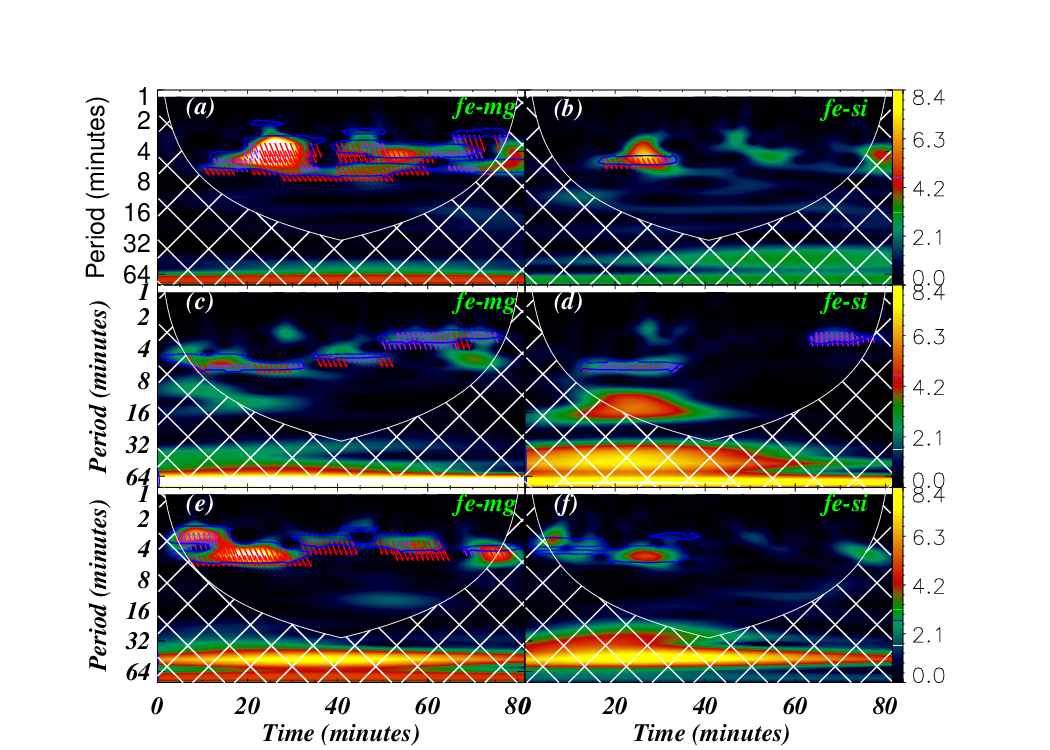} 
   	}
 \caption{The left panel displays cross power at various locations for the Fe-Mg pair, whereas the right panel displays the cross power for the Fe-Si pair. The remaining descriptions similar with those in Figure \ref{fig:figure6}.}
  \label{fig:figure10}
\end{figure*}

We calculated the cut-off frequency through phase analysis of velocity signals obtained from different heights at selected locations. This analysis provided the average cut-off frequency. We alos found several specific locations displayed diverse wave propagation patterns, as shown in Fig \ref{fig:figure10} for Fe-Mg and Fe-Si pairs. The right column shows the cross power between Fe-Mg pairs at various locations, while the left column displays cross power between Fe-Si pairs. Panels (a) and (b) indicated a dominance of downward propagating waves, while panels (c) and (d) demonstrated upward propagating waves. In panel (e), classical wave behavior can be seen, with shorter period waves propagating upward and longer period waves propagating downward. Panel (f) revealed no phase differences array on the cross wavelet power, indicating the absence of waves propagating up to the transition region at that specific pixel location.

\end{document}